\documentclass[11pt]{article}

\usepackage[margin=1in]{geometry}
\usepackage{amsmath,amssymb,amsthm}
\usepackage{bm}
\usepackage{graphicx}
\usepackage{booktabs}
\usepackage{array}
\usepackage{setspace}
\usepackage{url}
\usepackage{natbib}
\usepackage[colorlinks=true,citecolor=blue,linkcolor=blue,urlcolor=blue]{hyperref}
\usepackage{lineno}
\usepackage{tikz}
\usetikzlibrary{arrows.meta}
\usepackage{xspace}
\newcommand{\feat}[1]{\texttt{#1}\xspace}
\usepackage{indentfirst}

\title{Quaternion Nondecimated Wavelet Descriptors for Multiclass Breast Histology Classification}
\author{Sara Antonijevic and Brani Vidakovic\\
Department of Statistics, Texas A\&M University}
\date{\today}

\begin{document}
\maketitle

\begin{abstract}
Breast histology images carry diagnostic information in color, texture, orientation, and tissue architecture across a range of spatial scales. In H\&E microscopy this information is inherently chromatic and is not fully recovered when the red, green, and blue (RGB) channels are reduced to grayscale or transformed as independent scalar images. We propose an interpretable quaternion nondecimated wavelet framework for breast histology classification. Each RGB image is encoded as a pure quaternion field, and a quaternion nondecimated wavelet transform in two dimensions (QNDWT2D) produces multiscale, directional, color-coupled coefficient fields on the original image grid, keeping color as a single vector quantity rather than three separate channels. From these coefficients we build interpretable feature families summarizing stain balance, wavelet energy, amplitude heterogeneity, quaternion phase concentration, color-axis geometry, directional anisotropy, orientation entropy, and scale-dependent energy decay, each tied to a histopathological property such as nuclear density or glandular organization. We evaluate the descriptors on the BreAst Cancer Histology (BACH) challenge, a balanced four-class set of normal, benign, in situ, and invasive tissue, using a radial-kernel support vector machine (SVM) with repeated nested cross-validation. The descriptors yield balanced recognition across classes, with errors concentrated among adjacent categories while normal and invasive are rarely reversed. Permutation importance shows that directional, phase-concentration, anisotropy, scale, and amplitude-variability groups all contribute, indicating that the classifier draws on genuine quaternion and multiscale geometry rather than global color alone. The framework uses no pretrained networks, learned filters, or external databases, offering a reproducible, interpretable baseline for computational pathology.
\end{abstract}


\section{Introduction}
\label{sec:introduction}

Breast histopathology images contain diagnostic information at multiple spatial and chromatic scales. At fine scales, one observes nuclei, chromatin texture, cell boundaries, and local staining variation. At intermediate scales, one observes epithelial crowding, glandular distortion, ductal organization, and stromal interaction. At coarser scales, one observes broader tissue architecture and the spatial organization of normal, benign, in situ, and invasive patterns. The classification problem is therefore not only a problem of detecting isolated local objects. It is a problem of describing color, texture, orientation, and architecture jointly across scale. This is one reason that histopathology image analysis has become an important area of statistical imaging, machine learning, and computational pathology \citep{Gurcan2009,Veta2014,KomuraIshikawa2018}.

\begin{figure}[ht]
\centering
\begin{tabular}{cccc}
\includegraphics[width=0.22\textwidth]{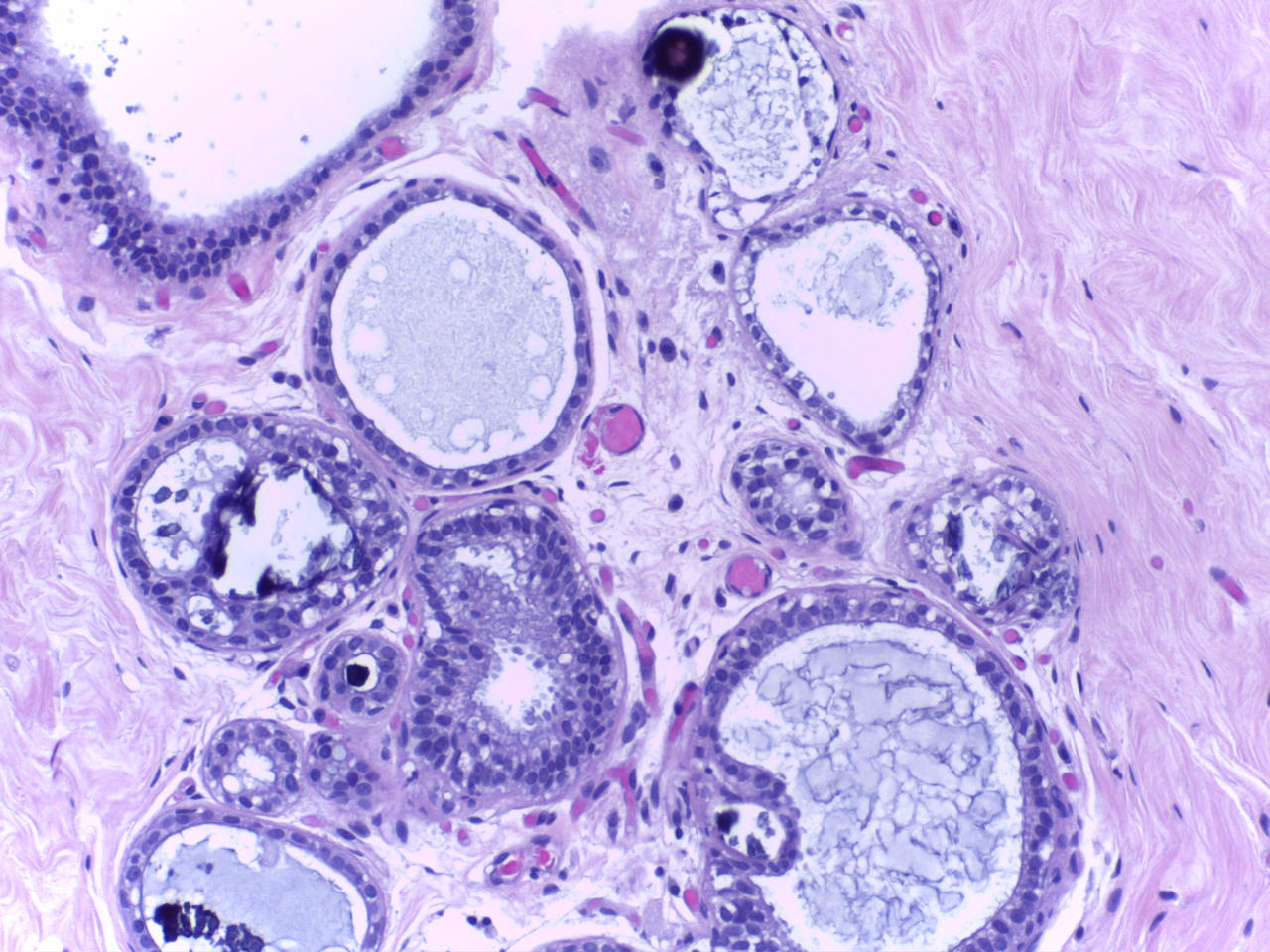} &
\includegraphics[width=0.22\textwidth]{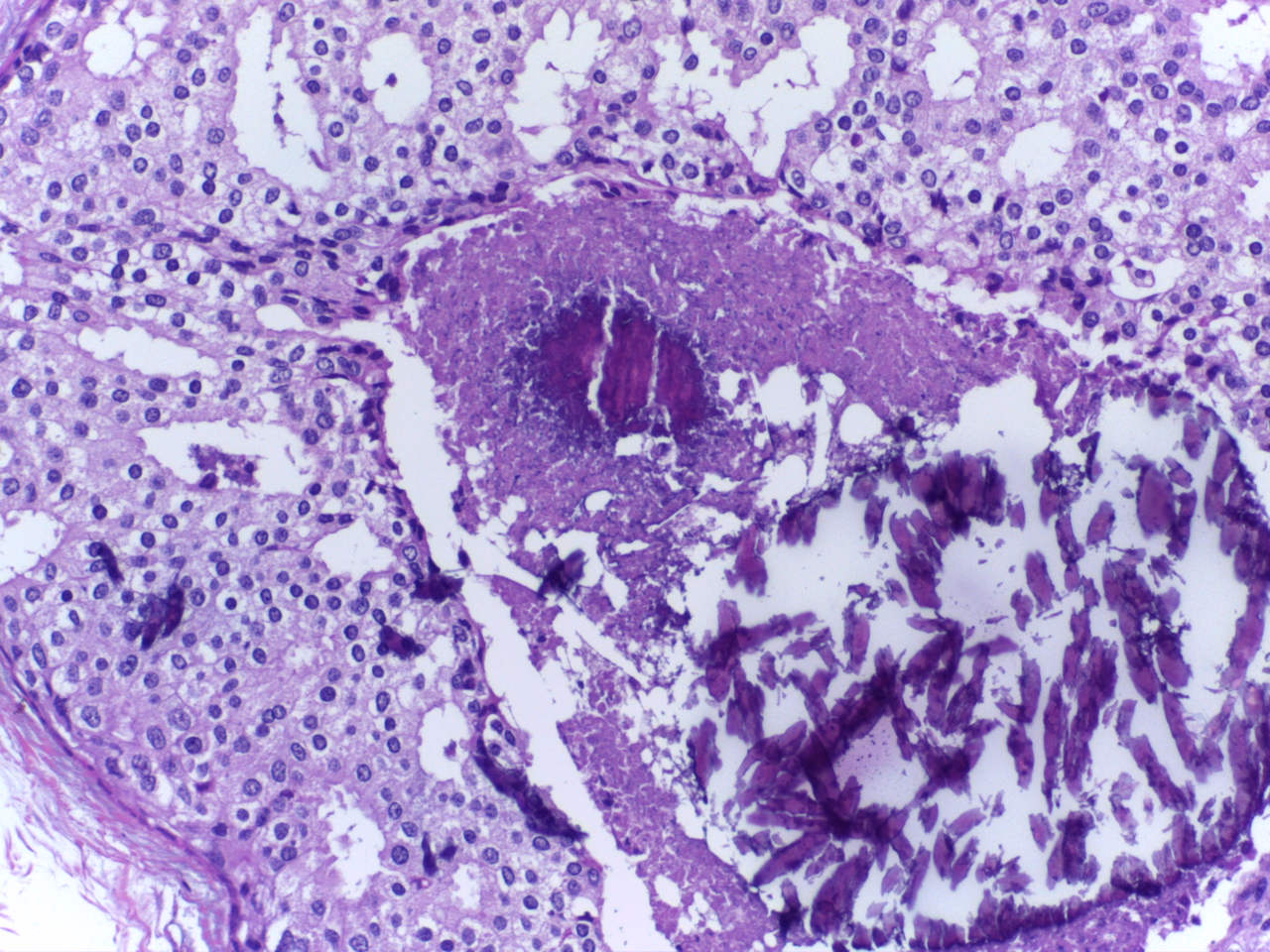} &
\includegraphics[width=0.22\textwidth]{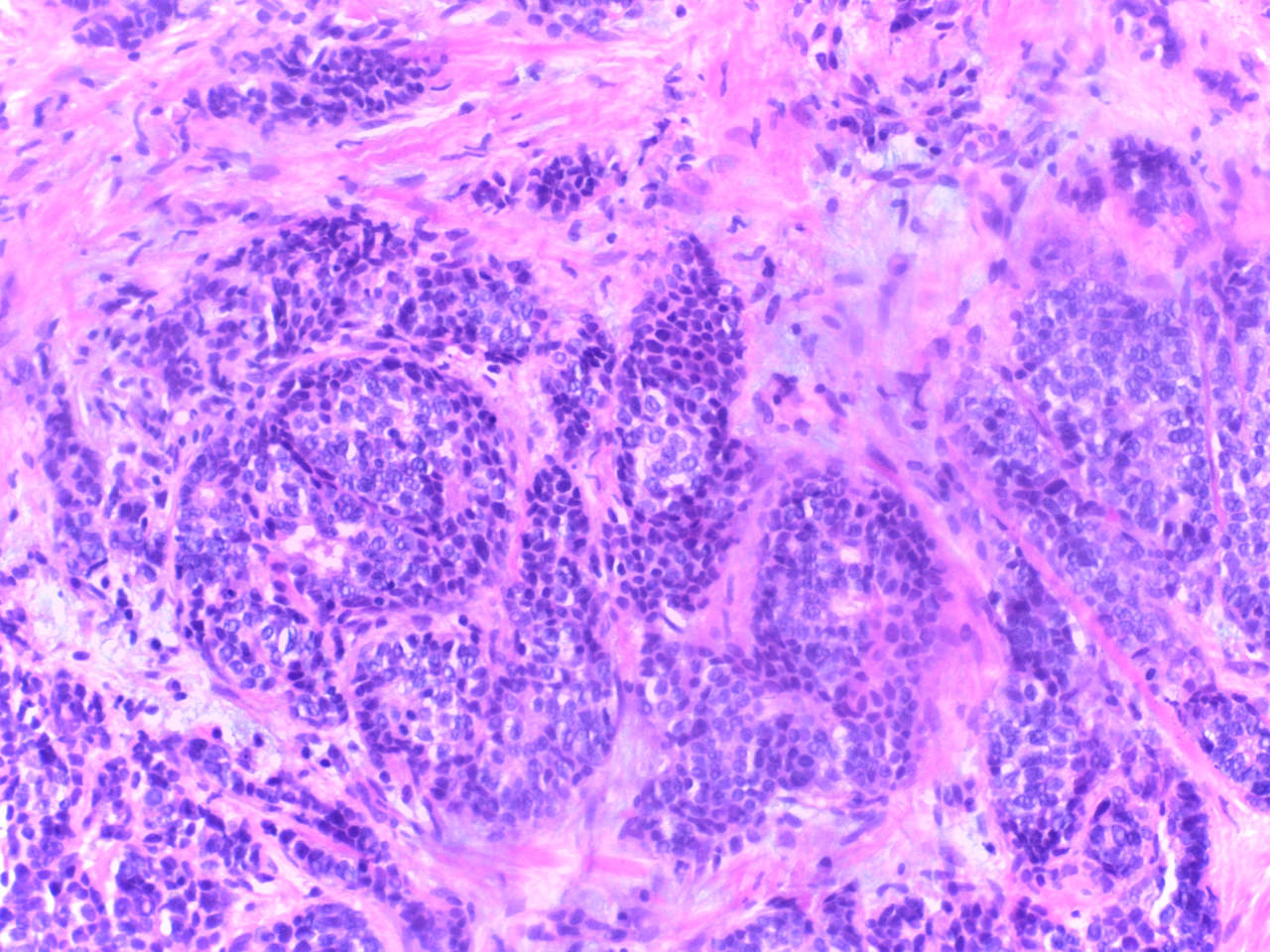} &
\includegraphics[width=0.22\textwidth]{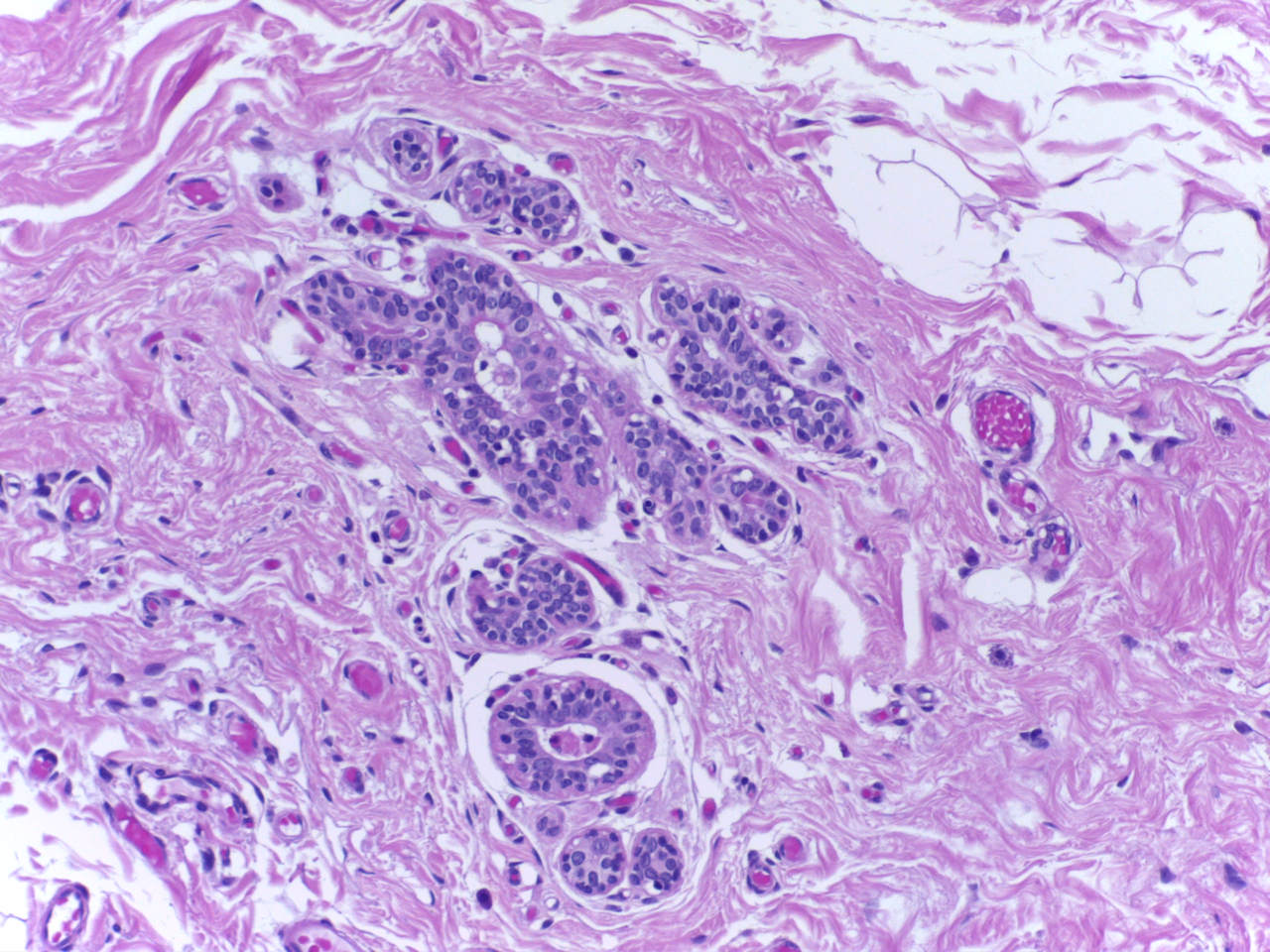} \\
Normal & Benign &   In situ & Invasive
\end{tabular}
\caption{Representative H\&E histology images from the four BACH diagnostic categories: normal, benign, in situ, and invasive. Color is preserved because it carries diagnostic information in H\&E stained tissue.}
\label{fig:mammo-four-categories-color}
\end{figure}

The BreAst Cancer Histology (BACH) challenge provides a natural testbed for this problem. It contains hematoxylin and eosin (H\&E) stained breast histology images from four diagnostic categories: normal tissue, benign lesion, in situ carcinoma, and invasive carcinoma \citep{aresta2019bach,bachchallenge2018dataset}. Figure~\ref{fig:mammo-four-categories-color} shows one representative image from each of the four classes. These classes differ through a combination of stain composition, nuclear density, glandular organization, stromal texture, and tissue architecture. Consequently, the dataset is well suited for evaluating image representations that attempt to preserve both color and multiscale spatial structure.

A large literature has demonstrated that deep learning methods perform well on breast histology classification. In the original BACH challenge, convolutional neural network (CNN) approaches were among the strongest methods \citep{aresta2019bach}, and subsequent studies have reported high classification rates using patch-based CNNs, transfer learning, and networks pretrained on large natural-image databases such as ImageNet \citep{Vesal2018,Nazeri2018}. These methods are effective classifiers, but the representation they learn is difficult to read in terms of identifiable histopathological mechanisms. A learned embedding may separate the classes well without revealing the roles of stain balance, directional organization, phase behavior, multiscale texture, or coarse architecture. This raises two questions: whether the four classes can be separated, and which image properties drive the separation. A learned embedding can settle the first without addressing the second.

This paper pursues that second question. We develop a quaternion nondecimated wavelet descriptor for color breast histology images, not as a replacement for deep networks but as a set of named, histopathologically meaningful measurements extracted before any classifier is fit. Each component of the descriptor has an intended interpretation: color balance, coefficient strength, amplitude heterogeneity, directional anisotropy, orientation entropy, quaternion phase behavior, color-vector geometry, and scale-dependent energy distribution. Classification then serves two purposes. It tests whether these interpretable quantities carry enough signal to support the four-class distinction, and through permutation importance it shows which of them the decision actually rests on. The aim is therefore not only to classify images, but to identify the multiscale color-textural information that matters for the classification.

A further methodological choice concerns color representation. A grayscale analysis collapses the RGB image into a single intensity channel before any multiscale step, discarding color information that is diagnostically relevant in H\&E stained tissue. A separate-channel RGB analysis transforms the red, green, and blue channels independently and combines their features only at the feature stage, which retains marginal color information but breaks the native coupling among channels. Quaternion coding keeps that coupling intact. Each RGB pixel is represented as a pure quaternion,
\begin{eqnarray}
A_0(x,y) &=& R(x,y)i + G(x,y)j + B(x,y)k,
\end{eqnarray}
so that the three channels are carried as one color vector rather than as three scalars. This idea is consistent with the broader literature on hypercomplex and quaternion representations for color image processing \citep{Bulow1999,Fletcher2018,FletcherSangwine2017}. The advantage for the present work is that cross-channel interaction is preserved through quaternion multiplication and reappears in the geometry of the coefficients, where it can be measured directly rather than reconstructed from separately transformed channels.

Quaternion wavelet methods have appeared in several forms, including quaternion analytic signals, dual-tree quaternion wavelets, quaternionic wavelets for texture classification, image denoising, and hypercomplex wavelet transforms of color-vector images \citep{BayroCorrochano2006,ChanChoiBaraniuk2008,SoulardCarre2011,GaiYangZhang2013,YinLiuShuiWu2012,Fletcher2018}. More recently, quaternion wavelets have been embedded as a preprocessing front end within learned architectures for medical image analysis, using a dual-tree quaternion wavelet transform whose sub-bands are packed into a quaternion and passed to a trained network \citep{SigilloGrassucciUnciniComminiello2025}. Across these forms the wavelet stage is decimated or dual-tree, and in the medical setting it typically feeds a trained network rather than serving as a fixed descriptor; such constructions also remain uncommon in applied histopathology. The contribution of this work is to pair quaternion color-vector coding with a two-dimensional nondecimated wavelet transform, a combination that, to our knowledge, has not been used in histopathology. It yields a redundant, shift-stable, multiscale representation that remains aligned with the original image grid while retaining RGB coupling directly in the quaternion coefficients.

We refer to this quaternion nondecimated wavelet transform in two dimensions as the QNDWT2D. Its coefficients are full quaternions, so each one carries several distinct kinds of information at once: a magnitude, a scalar-vector (phase-like) balance, a vector direction in the RGB-imaginary space, a directional subband, and a scale. Summarizing these across the image yields a feature family that is more interpretable than a generic learned embedding, yet richer than grayscale wavelet features or separate-channel RGB summaries.

The main contributions of the paper are as follows. First, we formulate a QNDWT2D representation for color histology images, encoding each image as a pure quaternion field so that the RGB channels are coupled at the representation stage rather than combined after feature extraction. Second, from the QNDWT2D domain we construct interpretable feature families with explicit histopathological meaning, covering energy, amplitude distribution, phase concentration, quaternion axis, directional anisotropy, orientation entropy, smooth-scale, and scale-slope summaries. Third, we test whether these descriptors carry enough signal for the four-class BACH microscopy problem, using a radial-kernel support vector machine (SVM) with repeated nested cross-validation. Fourth, we use permutation importance to identify which feature groups and individual predictors the classification decision rests on, linking discriminative power back to interpretable image properties.

The paper is organized as follows. Section~\ref{sec:Dataset} describes the BACH dataset and the four diagnostic classes. Section~\ref{sec:methodology} presents the QNDWT2D methodology and the two-sided operator convention used throughout. Section~\ref{sec:qndwt2-features} introduces the feature families and their histopathological motivation. Section~\ref{sec:results} reports the four-class classification results and the permutation-importance analysis. The supplementary material collects the supporting detail: quaternion algebra, quaternion filter construction, the left and right filtering conventions, the nondecimated synthesis formula, and the interpretation of quaternion wavelet coefficients. The concluding section returns to the main message, that QNDWT2D descriptors give an interpretable multiscale color-spatial representation for breast histology classification, complementary to high-performing but less transparent deep-learning approaches.

\section{Dataset}
\label{sec:Dataset}

The data used in this study come from the Grand Challenge on Breast Cancer Histology Images, organized in conjunction with the 2018 International Conference on Image Analysis and Recognition (ICIAR). The challenge was designed to evaluate automatic methods for classification and localization of clinically relevant breast histopathology patterns in microscopy images and whole-slide images. The microscopy component of the dataset contains H\&E stained breast tissue images classified into four diagnostic categories: normal tissue, benign lesion, in situ carcinoma, and invasive carcinoma \citep{aresta2019bach,bachchallenge2018dataset}.

The training portion of the microscopy dataset consists of 400 color images evenly distributed across the four diagnostic classes, 100 per class, so the task is balanced at the image level. Although the four labels follow an order of increasing diagnostic severity, we treat the task as four-class histopathological recognition rather than as a binary normal-versus-abnormal decision \citep{bachchallenge2018dataset}.

The BACH microscopy images are high-resolution RGB fields of size 2048 by 1536 pixels. They were acquired at 200$\times$ magnification, with a pixel size of approximately 0.42\,$\mu$m $\times$ 0.42\,$\mu$m. The images were collected during 2014, 2015, and 2017 from patients in the Porto and Castelo Branco regions of Portugal. The cases were provided by Ipatimup Diagnostics and originated from Hospital CUF Porto, Centro Hospitalar do T\^amega e Sousa, and Centro Hospitalar Cova da Beira. According to the BACH challenge description, image acquisition used a Leica DM 2000 LED microscope with a Leica ICC50 HD camera. The images were annotated by two expert pathologists from IPATIMUP and i3S \citep{aresta2019bach}.

The BACH dataset has become a standard benchmark for computational pathology studies of breast cancer histology image classification \citep{aresta2019bach}. It remains publicly available for research use through the challenge and archival repositories, subject to the stated license conditions \citep{polonia2019bachzenodo}.

\section{Methodology}
\label{sec:methodology}

The proposed analysis is based on a multiscale representation of the original color histology images. Two methodological choices are central. First, the transform should be applicable to images of arbitrary rectangular size, without requiring dyadic dimensions. Second, the transform should preserve the vector nature of RGB color, rather than reducing the image to grayscale or treating the three channels as unrelated scalar images.

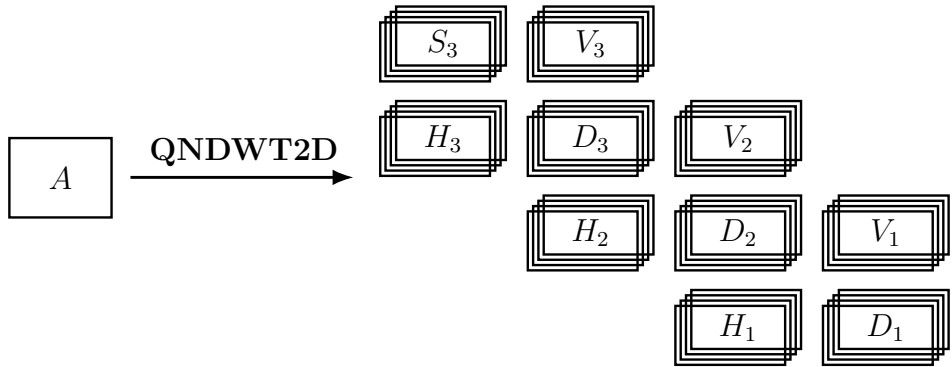
\begin{figure}[ht]
\centering
\begin{tikzpicture}[
    line width=0.9pt,
    >=Latex,
    font=\large
]

\def\w{1.45}
\def\h{0.78}

\def\dx{0.07}
\def\dy{0.07}

\newcommand{\qrect}[3]{%
    \draw (#1+3*\dx,#2+3*\dy) rectangle ++(\w,\h);
    \draw (#1+2*\dx,#2+2*\dy) rectangle ++(\w,\h);
    \draw (#1+\dx,#2+\dy) rectangle ++(\w,\h);
    \draw (#1,#2) rectangle ++(\w,\h);
    \node at (#1+0.5*\w+1.5*\dx,#2+0.5*\h+1.5*\dy) {#3};
}

\draw (-4.9,1.45) rectangle ++(1.35,1.05);
\node at (-4.225,1.975) {$A$};

\draw[->,line width=1.1pt] (-3.3,1.98) -- (-0.35,1.98);
\node[above] at (-1.8,1.98) {\textbf{QNDWT2D}};

\qrect{0.00}{3.25}{$S_3$}
\qrect{1.95}{3.25}{$V_3$}

\qrect{0.00}{2.00}{$H_3$}
\qrect{1.95}{2.00}{$D_3$}
\qrect{3.90}{2.00}{$V_2$}

\qrect{1.95}{0.75}{$H_2$}
\qrect{3.90}{0.75}{$D_2$}
\qrect{5.85}{0.75}{$V_1$}

\qrect{3.90}{-0.50}{$H_1$}
\qrect{5.85}{-0.50}{$D_1$}



\end{tikzpicture}
\caption{Schematic organization of the quaternion two-dimensional nondecimated wavelet transform. The input image $A$ is transformed by QNDWT2D into a hierarchy of same-size smooth and detail coefficient images. Each displayed subband on the right is shown as a close stack of four rectangles, representing the four quaternion-related outputs associated with that subband: the scalar part and the three imaginary components $i$, $j$, and $k$.}
\label{fig:qndwt_schematic}
\end{figure}

Classical two-dimensional discrete wavelet transforms apply recursive filtering followed by downsampling, which suits dyadic image sizes but forces padding, cropping, or resizing for arbitrary rectangular images and progressively changes the coefficient grid. We therefore use a two-dimensional nondecimated wavelet transform, which omits downsampling so that every scale component stays aligned with the original image coordinates and the method applies directly to rectangular histology images without imposing a dyadic structure.

The second choice concerns color. Because the red, green, and blue channels of an H\&E image are coupled rather than independent, grayscale conversion discards diagnostically meaningful color variation and separate-channel processing misses cross-channel relationships. A quaternion representation treats each RGB pixel as a single vector-valued object, following the literature on hypercomplex and quaternion image analysis \citep{Bulow1999,Fletcher2018,FletcherSangwine2017}.

\begin{figure}[ht]
\centering
\includegraphics[width=0.7\textwidth]{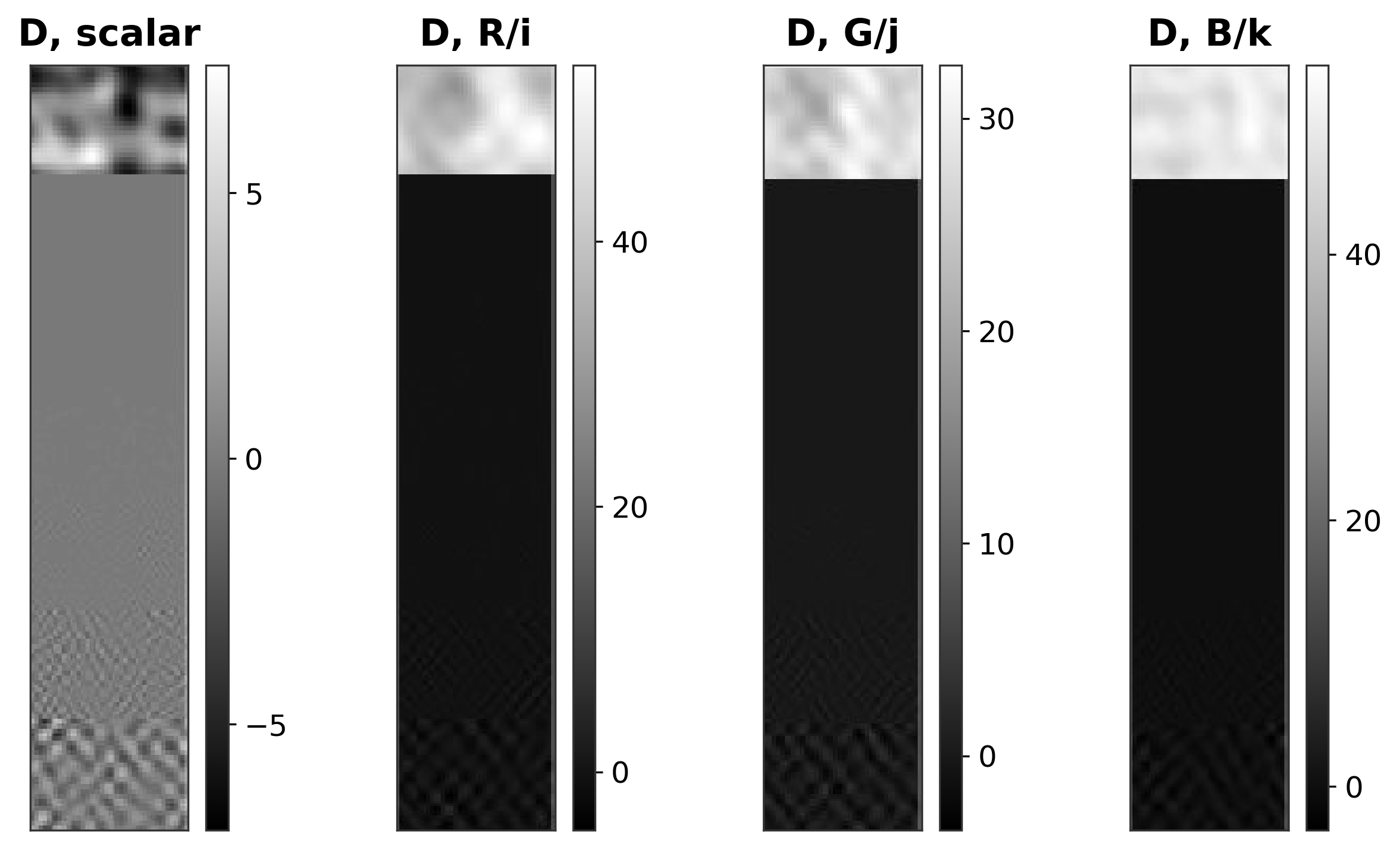}
\caption{Quaternion components in the diagonal hierarchy of the nondecimated wavelet transform. The four panels show the scalar component and the three imaginary color components, labeled as $D,\text{scalar}$; $D,R/i$; $D,G/j$; and $D,B/k$, respectively. The top rectangles are $S$ components. Here $D$ denotes the diagonal hierarchy of the nondecimated wavelet transform, while $R/i$, $G/j$, and $B/k$ denote the red, green, and blue channels represented as the $i$, $j$, and $k$ quaternion components. Grayscale bars show the component-specific coefficient ranges.}
\label{fig:quaternion-diagonal-ndwt-hierarchy}
\end{figure}

Let $R(x,y)$, $G(x,y)$, and $B(x,y)$ denote the normalized red, green, and blue intensities at pixel location $(x,y)$. We encode the image as a pure quaternion-valued field
\begin{eqnarray}
A_0(x,y) &=& R(x,y)i + G(x,y)j + B(x,y)k.
\end{eqnarray}
Thus the scalar part is initially zero, and the three observed color channels occupy the three imaginary quaternion components, so that subsequent filtering operations can mix and analyze the color components through quaternion algebra. Basic quaternion notation and algebraic identities are summarized in the supplementary material; only the notation needed for the transform is given here. Figure~\ref{fig:quaternion-diagonal-ndwt-hierarchy} shows the scalar part and the three imaginary color components of the diagonal detail subband for a representative image.

We describe the transform at the operator level. Let $\mathcal H_j$ denote the low-pass nondecimated quaternion convolution operator at level $j$, and $\mathcal G_j$ the corresponding high-pass operator. These symbols denote the full convolution operators induced by the level-$j$ dilated filters, not the short filter vectors. For an $m \times n$ image the left operator acts along rows and the right adjoint operator acts along columns, so the transform takes the standard two-sided form
\begin{eqnarray}
B_j &=& W_{1,j} A_{j-1} W_{2,j}^*,
\end{eqnarray}
where $W_{1,j}$ and $W_{2,j}$ each denote either $\mathcal H_j$ or $\mathcal G_j$, according to the subband. In implementation the large matrices are not formed explicitly; their action is computed by circular shifts and left or right quaternion multiplication by filter taps. Figure~\ref{fig:qndwt_schematic} summarizes the multilevel organization of the smooth and detail subbands produced by the transform. At level $j$, starting from the current approximation image $A_{j-1}$, the transform produces one smooth component and three directional detail components:
\begin{eqnarray}
S_j &=& \mathcal H_j A_{j-1} \mathcal H_j^*, \\
H_j &=& \mathcal G_j A_{j-1} \mathcal H_j^*, \\
V_j &=& \mathcal H_j A_{j-1} \mathcal G_j^*, \\
D_j &=& \mathcal G_j A_{j-1} \mathcal G_j^*.
\end{eqnarray}
Here $S_j$ is the smooth component. The detail image $H_j$ is high-pass in the row direction and low-pass in the column direction, $V_j$ is low-pass in the row direction and high-pass in the column direction, and $D_j$ is high-pass in both directions. The approximation is updated recursively by
\begin{eqnarray}
A_j &=& S_j,
\end{eqnarray}
and the procedure is repeated for the desired number of levels. In this work we use $L = 6$ levels for all images.

Because the transform is nondecimated, each of $S_j$, $H_j$, $V_j$, and $D_j$ has the same spatial dimensions as the original image, so multiscale information is retained on the original image lattice rather than being successively reduced to smaller grids.

This construction combines two useful properties. The nondecimated structure gives a redundant, shift-stable, multiscale representation on the original image lattice, and the quaternion structure lets the transform capture joint spatial and chromatic variation. Related quaternion wavelet ideas have been used in coherent multiscale image processing, texture classification, image denoising, and color-vector image processing \citep{BayroCorrochano2006,ChanChoiBaraniuk2008,SoulardCarre2011,GaiYangZhang2013,YinLiuShuiWu2012,Fletcher2018}. In the present work, this transform serves as the image representation from which diagnostic summaries are subsequently computed. The precise filter construction, quaternion multiplication order, dilation convention, and reconstruction formula are given in the supplementary material.

\section{Feature Families in the QNDWT2D Domain}
\label{sec:qndwt2-features}
The QNDWT2D representation produces a multiscale collection of quaternion-valued coefficient fields. At each level $j$, the transform gives one smooth component $S_j$ and three directional detail components $H_j$, $V_j$, and $D_j$, defined by the two-sided operator construction of Section~\ref{sec:methodology}. The $H$, $V$, and $D$ feature labels below follow the operator definitions in Section~\ref{sec:methodology} and the supplementary material.

For orientation $O \in \{H,V,D\}$ and level $j$, let $C_{O,j}(x,y)$ denote a quaternion coefficient,
\begin{eqnarray}
C_{O,j}(x,y)
&=&
a_{O,j}(x,y)
+
b_{O,j}(x,y)i
+
c_{O,j}(x,y)j
+
d_{O,j}(x,y)k.
\end{eqnarray}
Its magnitude is
\begin{eqnarray}
\rho_{O,j}(x,y)
&=&
|C_{O,j}(x,y)| \nonumber\\
&=&
\sqrt{
a_{O,j}^2(x,y)
+
b_{O,j}^2(x,y)
+
c_{O,j}^2(x,y)
+
d_{O,j}^2(x,y)
}.
\end{eqnarray}
The features are constructed by summarizing the coefficient fields across pixels, orientations, and scales. The purpose is not to retain the full transform, but to form an image-level descriptor that captures the dominant multiscale, directional, and color-geometric behavior of the histology image.

\paragraph{RGB marginal summaries.}
The first group consists of simple summaries of the original RGB image: channel means, channel standard deviations, and pairwise channel correlations. These features do not use the wavelet transform directly. They describe global stain intensity, stain variability, and overall association among the red, green, and blue channels. These summaries are included because histology images can differ substantially in stain balance and stain variability. Although the central methodology is based on quaternion wavelet coefficients, it is useful to retain a small set of direct color summaries. They provide a baseline description of the image before multiscale analysis and help distinguish purely global color effects from texture and orientation effects.

\paragraph{Energy and log-energy summaries.}
The second group consists of energy and log-energy summaries for each directional subband and level. For orientation $O$ and level $j$, the energy is the average squared quaternion magnitude,
\begin{eqnarray}
E_{O,j}
&=&
{1 \over N}
\sum_{x,y}
|C_{O,j}(x,y)|^2,
\end{eqnarray}
where $N$ is the number of pixels. The corresponding log-energy is computed on the stabilized energy scale. These features measure the amount of wavelet-domain activity at a specific scale and direction. In breast histology, diagnostic classes may differ in nuclear density, glandular structure, stromal organization, epithelial crowding, and boundary irregularity. Such differences are naturally reflected in directional multiscale activity. Fine levels tend to respond to small structures such as nuclei and sharp boundaries, while coarser levels respond to broader tissue organization. Log-energy is included because wavelet energies often vary over a wide numerical range and are more stable on a logarithmic scale.

\paragraph{Amplitude distribution summaries.}
Energy measures the average squared magnitude, but it does not describe the full distribution of coefficient amplitudes. Therefore, the third group summarizes the empirical distribution of $\rho_{O,j}(x,y)$ within each subband. This family includes measures of central tendency, spread, asymmetry, tail behavior, and histogram entropy. The motivation is that two images may have similar total subband energy but very different coefficient distributions. A subband response may be diffuse across the tissue field, concentrated in a few high-activity regions, or dominated by isolated boundaries or dense cellular areas. Distributional summaries help distinguish these cases. Skewness and kurtosis are especially useful for detecting sparse strong responses, while entropy describes whether the subband amplitudes are concentrated or broadly dispersed.
\paragraph{Quaternion phase summaries.}
A quaternion coefficient contains both scalar and vector information. We write
\begin{eqnarray}
C_{O,j}(x,y)
&=&
{\rm Sc}\{C_{O,j}(x,y)\}
+
{\rm Vec}\{C_{O,j}(x,y)\}.
\end{eqnarray}
A phase-like angle is defined from the relative size of the vector and scalar parts,
\begin{eqnarray}
\theta_{O,j}(x,y)
&=&
\arctan
\left(
{|{\rm Vec}\{C_{O,j}(x,y)\}|
\over
{\rm Sc}\{C_{O,j}(x,y)\}}
\right),
\end{eqnarray}
computed with the two-argument arctangent that respects the sign of the scalar part. Because $|{\rm Vec}\{C_{O,j}(x,y)\}| \ge 0$, the angle lies in $[0,\pi]$. The corresponding feature family summarizes the circular behavior of these phase angles. In particular, phase concentration describes whether the scalar-vector balance of the quaternion coefficients is coherent within a subband or highly dispersed. These summaries are included because quaternion wavelet coefficients carry more information than magnitude alone. The scalar-vector balance reflects the local color-spatial response induced by quaternion filtering, and its variability may be associated with heterogeneous staining, irregular morphology, or complex local tissue structure.

\paragraph{Quaternion axis summaries.}
For coefficients with nonzero vector part, the normalized vector direction is
\begin{eqnarray}
u_{O,j}(x,y)
&=&
{{\rm Vec}\{C_{O,j}(x,y)\}
\over
|{\rm Vec}\{C_{O,j}(x,y)\}|}.
\end{eqnarray}
This direction lies in the three-dimensional imaginary quaternion space spanned by $i$, $j$, and $k$, the same space used to encode the RGB channels of the original image. Axis summaries describe the dominant direction and concentration of the quaternion vector parts. Their role is different from amplitude summaries. Amplitude describes how strong the local response is, while the axis describes how the RGB components participate in that response. This distinction is important in color histology because two tissue regions may have comparable wavelet magnitude but different stain composition or different color-vector geometry.

\paragraph{Directional balance and anisotropy summaries.}
At each level, the three directional energies are converted into relative shares,
\begin{eqnarray}
p_{H,j}
&=&
{E_{H,j}
\over
E_{H,j}+E_{V,j}+E_{D,j}},\\
p_{V,j}
&=&
{E_{V,j}
\over
E_{H,j}+E_{V,j}+E_{D,j}},\\
p_{D,j}
&=&
{E_{D,j}
\over
E_{H,j}+E_{V,j}+E_{D,j}},
\end{eqnarray}
The horizontal-vertical anisotropy and the diagonal share are
\begin{eqnarray}
A_{HV,j} &=& {E_{H,j}-E_{V,j} \over E_{H,j}+E_{V,j}}, \\
A_{D,j}  &=& {E_{D,j} \over E_{H,j}+E_{V,j}+E_{D,j}} = p_{D,j}.
\end{eqnarray}
The signed quantity $A_{HV,j}$ measures imbalance between row-directional
and column-directional activity, and $A_{D,j}$ measures the relative
contribution of diagonal activity. Note that $A_{D,j}$ and $p_{D,j}$ are
the same quantity under two names, which is why only one of them is
retained as a nonredundant predictor in
Section~\ref{subsec:individual-importance}. These features are motivated by the directional organization of tissue. Glandular boundaries, stromal fibers, epithelial interfaces, and invasive growth patterns can produce directional preferences. Relative directional summaries reduce dependence on total image activity and focus on how activity is distributed across orientations. They therefore capture whether the image response is predominantly row-directional, column-directional, diagonal, or more balanced.

\paragraph{Orientation entropy.}
Orientation entropy summarizes the distribution of energy across $H$, $V$, and $D$ at a given level. Using the directional shares, it has the form
\begin{eqnarray}
{\cal E}_j
&=&
-
{1 \over \log 3}
\sum_{O \in \{H,V,D\}}
p_{O,j}\log p_{O,j}.
\end{eqnarray}
The value is small when the energy is concentrated in one orientation and large when energy is more evenly distributed across orientations. This feature is intended to measure directional organization. A more orderly or strongly oriented tissue pattern may yield lower orientation entropy, while a mixed or irregular tissue structure may yield a more balanced directional response. In this way, orientation entropy complements the anisotropy features by describing the overall directional dispersion.

\paragraph{Smooth-scale and global energy summaries.}
The final smooth component and the energy summaries aggregated across levels form another feature group. These descriptors include the energy of the final smooth component, total detail energy, directional total energies, and average smooth energy across levels. These features are included because not all diagnostic differences appear only as local detail. Some differences are expressed through broad tissue organization, large-scale staining patterns, or coarse architectural structure. The smooth component captures low-frequency content after repeated filtering, while aggregated energy summaries describe the total multiscale activity in each direction.

\paragraph{Scale-distribution and slope summaries.}
The level-wise energy sequences are also summarized as scale distributions. For a fixed orientation $O$, the sequence
\begin{eqnarray}
E_{O,1}, E_{O,2}, \ldots, E_{O,L}
\end{eqnarray}
describes how activity changes across resolution. We summarize this behavior using spectral entropy across levels and by fitting a simple log-energy slope,
\begin{eqnarray}
\log_2 E_{O,j}
&=&
\alpha_O + \beta_O j + \epsilon_{O,j}.
\end{eqnarray}
The slope $\beta_O$ gives a compact measure of how rapidly detail energy changes with scale. These summaries are motivated by the multiscale nature of histology. Normal, benign, in situ, and invasive tissue may differ not only in the amount of activity, but also in how that activity is distributed from fine nuclear texture to coarser architecture. Scale-distribution features are therefore intended to capture roughness, persistence of structure across levels, and the relative dominance of fine or coarse patterns.

\subsection{Rationale for the complete feature family}
\label{subsec:feature-rationale}

The QNDWT2D descriptor is intended to be interpretable rather than merely high-dimensional. Each group of features corresponds to a recognizable aspect of histopathological image content: stain balance, local activity, texture heterogeneity, orientation, coarse architecture, and multiscale organization. This is important because breast histology images are not characterized by a single visual attribute. Diagnostic appearance is produced jointly by nuclear density, chromatin texture, epithelial organization, glandular or ductal structure, stromal interaction, boundary irregularity, and stain composition. Reviews of histopathology image analysis emphasize precisely these types of measurements, including color, texture, nuclear morphology, spatial organization, and multiresolution image representations \citep{Gurcan2009,Veta2014,KomuraIshikawa2018}.

The RGB marginal features have a direct stain-related interpretation. H\&E images carry diagnostic information through color composition, since nuclear-rich regions, cytoplasm, stroma, and extracellular material differ in stain response. This is one reason color normalization, stain separation, and color deconvolution have been important in digital pathology \citep{RuifrokJohnston2001,Veta2014}. In our setting, the RGB means, standard deviations, and correlations provide a low-dimensional summary of global stain balance and cross-channel association. These features do not attempt to describe morphology, but they record whether an image differs in overall color content, staining variability, or channel coupling before any multiscale decomposition is considered.

The energy and log-energy features measure the amount of activity in each directional subband and at each scale. Histopathologically, these features are related to the density and strength of local transitions. Fine-scale energy is expected to respond to nuclei, chromatin variation, small edges, and local stain granularity. Intermediate-scale energy reflects larger cellular aggregates, glandular boundaries, ductal structures, and local epithelial-stromal organization. Coarser-scale energy reflects broader tissue architecture and low-frequency tissue arrangement. Because the QNDWT2D is nondecimated, these features remain defined on the original image lattice while still probing progressively coarser structure at higher levels.

Amplitude distribution features complement energy features. The mean and standard deviation of quaternion coefficient magnitudes measure typical response and heterogeneity. Skewness and kurtosis describe asymmetry and tail behavior, and are useful when a subband is dominated by localized structures such as dense nuclei, sharp glandular edges, necrotic or stromal boundaries, or irregular invasive patterns. Entropy of the amplitude distribution describes whether subband activity is concentrated or broadly distributed. These are wavelet-domain analogues of classical texture measurements, and texture features have long been used in image classification and histopathology analysis \citep{Haralick1973,Gurcan2009,Veta2014}.

The directional balance and anisotropy features have a particularly natural histopathological interpretation. Tissue often has preferred local orientations, and the relative directional shares $p_{H,j}$, $p_{V,j}$, and $p_{D,j}$ describe how much of the level-$j$ activity is organized in the three wavelet directions. The horizontal-vertical contrast $A_{HV,j}$ measures imbalance between row-directional and column-directional activity, while the diagonal share $A_{D,j}$ measures the contribution of oblique or corner-like structure. Such orientation information is not merely mathematical. Quantitative nuclear shape and orientation features extracted from H\&E breast histology have been shown to have prognostic value, supporting the broader idea that orientation and alignment carry biologically meaningful information \citep{Lu2018}. In our transform, orientation is measured at several spatial scales, so the features can reflect whether anisotropy is a fine nuclear phenomenon, an intermediate glandular phenomenon, or a coarse architectural phenomenon.

Orientation entropy summarizes whether directional activity is concentrated or mixed. A low value indicates energy concentrated in one direction and a high value indicates energy spread across directions. In breast histology this distinction is potentially meaningful. Normal or benign tissue may preserve more organized ductal or stromal patterns, whereas malignant or invasive tissue may show more irregular architecture. Orientation entropy is therefore a compact measure of directional disorder, complementary to the signed anisotropy measures.

Smooth-scale and global energy features capture the low-frequency component of the image. Not all diagnostically relevant information is edge-like or high-frequency. Broad tissue composition, large glandular fields, stromal expansion, and coarse epithelial organization can appear in the smooth component or in aggregate detail-energy summaries across levels. The final smooth energy and total directional energies therefore connect the descriptor to coarse architecture, while the level-specific details connect it to finer texture and morphology.

Scale-distribution and slope features summarize how activity is allocated across resolution levels. In a nondecimated transform, level $j$ is obtained by inserting $2^{j-1}-1$ zeros between filter taps. This means that the filter support expands by a factor of approximately two from one level to the next. Although the coefficient arrays remain the same image size, the frequency band and spatial scale being probed become progressively coarser. The spectral entropy across levels measures whether activity is concentrated at a few scales or distributed broadly over scales. The log-energy slope measures whether detail energy decays rapidly or slowly with scale. Histopathologically, these features are meant to distinguish images dominated by fine nuclear texture from images with persistent coarse architectural organization. Wavelet-based multiresolution texture analysis has been used in histological image studies, including breast cancer tissue, precisely because tissue appearance depends on scale \citep{Hwang2005,Chaddad2018}.

The quaternion phase and axis features are less standard in classical histopathology, but they matter here because the image is not transformed as a grayscale scalar field. The phase summaries capture the coherence or dispersion of the scalar-vector balance induced by quaternion filtering, and the axis summaries capture how the RGB components combine in the local wavelet response. These features are motivated by the fact that color histology is not simply intensity texture. H\&E staining creates coupled color-spatial patterns, and quaternion filtering lets these interactions appear directly in the coefficient geometry rather than being reconstructed later from separate per-channel transforms.

Taken together, the feature family is designed to cover the main visual mechanisms by which histology classes differ. RGB summaries describe global stain behavior. Energy and log-energy describe the strength of multiscale activity. Amplitude distribution summaries describe heterogeneity, sparsity, and tail behavior. Directional shares, anisotropy, and orientation entropy describe tissue alignment and directional disorder. Smooth-scale and slope features describe coarse architecture and the scale distribution of activity. Quaternion phase and axis summaries describe color-vector geometry and scalar-vector coupling. Thus the QNDWT2D descriptor is not a collection of arbitrary numerical features. It is a structured image phenotype: color, texture, orientation, scale, and quaternion color geometry are all measured in a common nondecimated multiscale representation.

\section{Classification Results}
\label{sec:results}

This section reports the classification results obtained with the QNDWT2D transform of Section~\ref{sec:methodology}, using the descriptor families defined in Section~\ref{sec:qndwt2-features}.

\subsection{Radial SVM classifier}
\label{subsec:svm-method}
The radial SVM was selected as the primary classifier because it matched the structure of the descriptor set. The QNDWT2D feature table is moderate-dimensional, nonlinear, and highly correlated by construction: several feature families summarize the same coefficient fields from different perspectives, including energy, amplitude distribution, phase, axis, direction, and scale. A linear sparse method such as an elastic-net model is attractive for variable screening, but it imposes an essentially additive linear decision rule on a feature space whose discriminating information is expected to involve nonlinear interactions among scale, orientation, stain behavior, and quaternion phase-axis geometry. AdaBoost and related boosting approaches can also be useful competitors, but in this setting they tend to be more sensitive to small-sample fold-to-fold variation and to the presence of many correlated summaries. The radial SVM offers a useful compromise: it allows nonlinear decision boundaries, has only a small number of tuning parameters, is well suited to standardized continuous predictors, and can be evaluated cleanly inside a nested cross-validation scheme.

Classification was performed using a support vector machine with a radial basis kernel. The four diagnostic classes, Normal, Benign, InSitu, and Invasive, were handled through an error-correcting output-code formulation with one-versus-one binary learners. Predictors were standardized within the training folds, and all model tuning was carried out on the training data only.

The reported results are based on repeated nested cross-validation. The outer loop used 100 repetitions of 5-fold cross-validation, giving 500 held-out folds, and the inner loop used 5-fold cross-validation to tune the box constraint and the radial kernel scale. In each outer fold, 303 nonconstant predictors were retained. Because the dataset contains 400 images and the outer procedure was repeated 100 times, the aggregated confusion table contains 40{,}000 held-out predictions, with each true class represented 10{,}000 times. The repetitions reuse the same 400 images, so they are not independent and the fold-level standard deviations reported below understate the true sampling variability; the aggregate is best read as a stable description of the classifier's error structure rather than as an independent test estimate.

\subsection{Classification performance}
\label{subsec:svm-confusion}
Across the 500 outer folds, the mean overall accuracy was 0.686, the mean balanced accuracy was 0.686, and the mean macro F$1$ score (the unweighted mean of the per-class F$1$ scores, each the harmonic mean of precision and recall) was 0.685, with standard deviations of approximately 0.048 for all three measures. These are averages of the per-fold metrics; recomputed from the pooled confusion table in Table~\ref{tab:svm-confusion},the macro F$1$ is 0.687; the small difference reflects the gap between averaging per-fold scores and pooling. The classifier therefore improves clearly over the 25\% chance level for a balanced four-class problem, but the task remains difficult, as expected for a histopathology problem with adjacent diagnostic categories.

\begin{table}[!htbp]
\centering
\caption{\small Repeated cross-validation confusion table for the radial SVM. Percentages are conditional on the true class, so each column sums to 100\%.}
\label{tab:svm-confusion}
\vspace*{0.15in}
\begin{tabular}{lrrrr}
\hline
 & True Normal & True Benign & True InSitu & True Invasive\\
\hline
Pred Normal   & 6862 (68.6\%) & 1086 (10.9\%) &  898 (9.0\%)  &  299 (3.0\%)\\
Pred Benign   & 1632 (16.3\%) & 6945 (69.5\%) & 1286 (12.9\%) & 1656 (16.6\%)\\
Pred InSitu   & 1231 (12.3\%) &  925 (9.2\%)  & 7084 (70.8\%) & 1513 (15.1\%)\\
Pred Invasive &  275 (2.8\%)  & 1044 (10.4\%) &  732 (7.3\%)  & 6532 (65.3\%)\\
\hline
\end{tabular}
\end{table}

Table~\ref{tab:svm-confusion} gives the aggregated confusion table, with rows representing predicted classes and columns representing true classes, so the diagonal entries are the class-wise recalls. The recalls are similar across classes, at 68.6\% for Normal, 69.5\% for Benign, 70.8\% for InSitu, and 65.3\% for Invasive. The classifier therefore does not simply favor one dominant class but achieves a relatively balanced recognition rate across the four diagnostic groups.

The off-diagonal structure is also informative. True Normal images are most often confused with Benign (16.3\%), followed by InSitu (12.3\%). When misclassified, true Benign images are spread across the remaining classes, at 10.9\% Normal, 9.2\% InSitu, and 10.4\% Invasive. True InSitu images are most often confused with Benign (12.9\%), followed by Normal (9.0\%) and Invasive (7.3\%). True Invasive images are most often confused with Benign and InSitu (16.6\% and 15.1\%) and are only rarely confused with Normal.

This pattern is histopathologically plausible, since the most frequent errors occur among categories that can share local texture, staining, or intermediate architectural features. Direct Normal-Invasive reversals remain rare: 2.8\% of true Normal images are predicted as Invasive, and 3.0\% of true Invasive images are predicted as Normal. As shown in Table~\ref{tab:svm-class-performance}, the Invasive class has the highest precision (0.761) and Normal is close behind (0.750), whereas Benign has the lowest precision (0.603) because it attracts false positives from all other classes. In this sense, Benign functions as a broad intermediate attractor in the current feature space. 

\begin{table}[!htbp]
\centering
\caption{\small Class-wise performance of the radial SVM, aggregated over repeated cross-validation.}
\label{tab:svm-class-performance}
\vspace*{0.15in}
\begin{tabular}{lrrr}
\hline
Class & Recall & Precision & F1\\
\hline
Normal   & 0.686 & 0.750 & 0.717\\
Benign   & 0.695 & 0.603 & 0.645\\
InSitu   & 0.708 & 0.659 & 0.683\\
Invasive & 0.653 & 0.761 & 0.703\\
\hline
\end{tabular}
\end{table}

\begin{figure}[!htbp]
\centering
\includegraphics[width=0.7\textwidth]{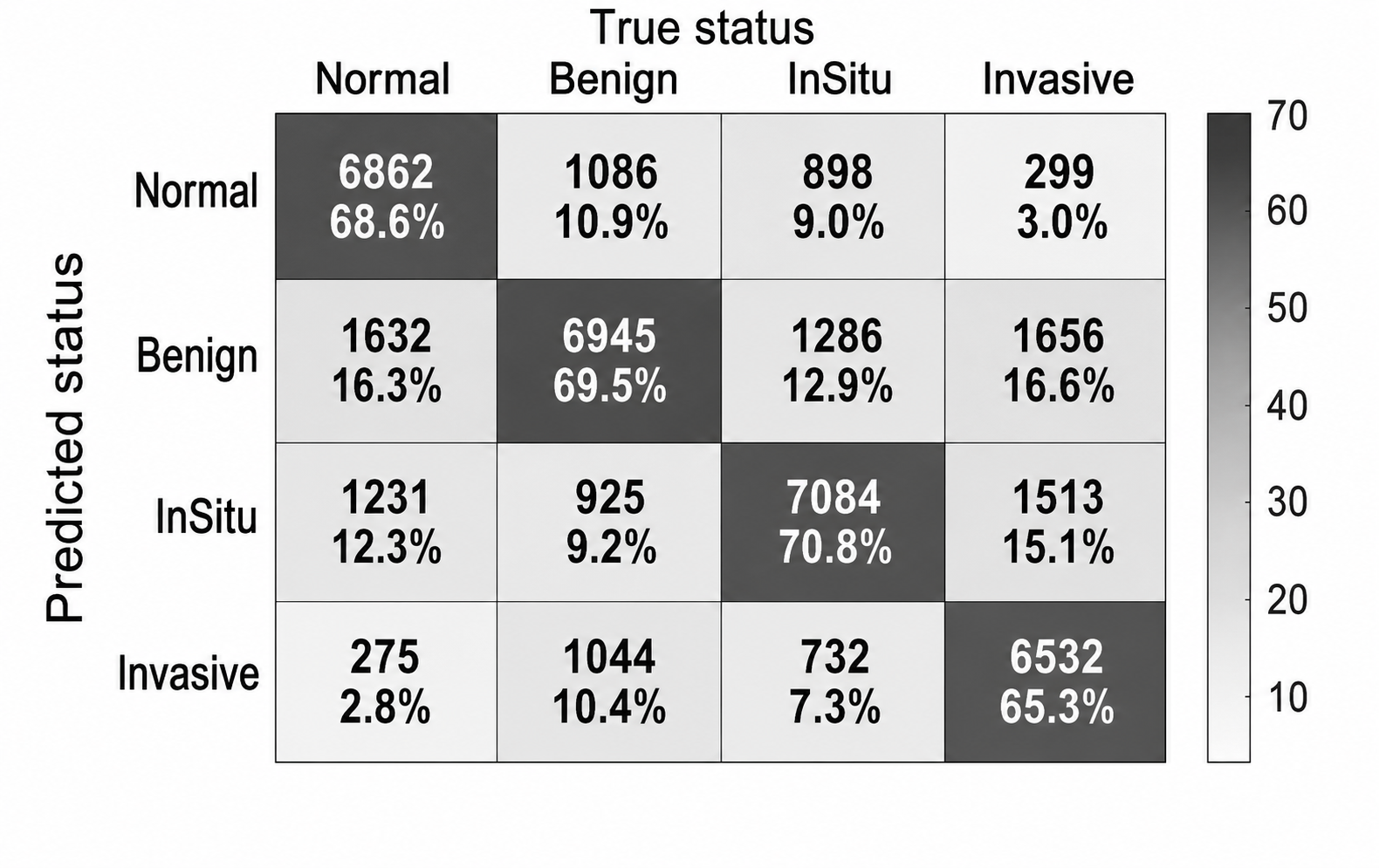}
\caption{\small Radial SVM confusion display. Rows are predicted classes and columns are true classes. Each cell shows the repeated cross-validation count and the corresponding percentage conditional on the true class.}
\label{fig:svm-confusion}
\end{figure}

\subsection{Marginal binary summaries from the four-class classifier}
\label{subsec:marginal-binary-summaries}

Although the primary analysis is a four-class problem, several marginal binary
contrasts derived from the aggregated four-class confusion table are also
informative. Throughout this subsection the numeric labels $0$, $1$, $2$, and
$3$ denote Normal, Benign, InSitu, and Invasive, respectively. These summaries do
not fit new binary classifiers. They describe how the already fitted four-class
radial SVM separates selected pairs or groups of diagnoses, and we report each
pairwise contrast in two ways so that the effect of the off-pair predictions is
visible directly.

For a pairwise contrast, some images whose true label belongs to the pair are
assigned by the four-class classifier to a category outside the pair. In the $0$
versus $3$ contrast, for example, an image whose true label is Normal or Invasive
may be predicted as Benign or InSitu. We call these off-pair predictions. The
conditional summary excludes them and is computed only on images the classifier
assigned to one of the two contrasted classes. The unconditional summary retains
them and counts every off-pair prediction as an error, so that the classifier is
required to produce a within-pair decision for every relevant image. The grouped
contrast $\lbrace 0,1\rbrace$ versus $\lbrace 2,3\rbrace$ has no off-pair set,
because every class belongs to one of the two groups, so its conditional and
unconditional summaries coincide.

\begin{table}[!htbp]
\centering
\caption{\small Conditional and unconditional marginal binary performance derived
from the aggregated four-class radial SVM predictions. Conditional summaries
exclude off-pair predictions; unconditional summaries count every off-pair
prediction as an error, so the classifier must decide for every relevant image.
For $0$ vs.\ $3$, class $3$ is positive; for $1$ vs.\ $2$, class $2$ is positive;
for $1$ vs.\ $3$, class $3$ is positive; and for $\lbrace 0,1\rbrace$ vs.\
$\lbrace 2,3\rbrace$, the positive group is $\lbrace 2,3\rbrace$. All metric
entries are percentages. The Used and Off-pair counts apply to both treatments.
PPV and NPV are identical across treatments because off-pair predictions belong to
neither contrasted class; accuracy, sensitivity, specificity, and F1 differ. The grouped contrast has no off-pair predictions, so its conditional and
unconditional summaries coincide and are shown on a single line marked ``Both''.}
\label{tab:marginal-binary-conditional}
\vspace*{0.15in}
\begin{tabular}{llrrrrrrrr}
\hline
Contrast & Treat. & Acc. & SE & SP & PPV & NPV & F1 & Used & Off-pair\\
\hline
$0$ vs.\ $3$ & Cond.   & 95.9 & 95.6 & 96.1 & 96.0 & 95.8 & 95.8 & 13{,}968 & 6{,}032\\
             & Uncond. & 67.0 & 65.3 & 68.6 & 96.0 & 95.8 & 77.7 &          &        \\
\hline
$1$ vs.\ $2$ & Cond.   & 86.4 & 84.6 & 88.2 & 88.5 & 84.4 & 86.5 & 16{,}240 & 3{,}760\\
             & Uncond. & 70.1 & 70.8 & 69.5 & 88.5 & 84.4 & 78.6 &          &        \\
\hline
$1$ vs.\ $3$ & Cond.   & 83.3 & 79.8 & 86.9 & 86.2 & 80.7 & 82.9 & 16{,}177 & 3{,}823\\
             & Uncond. & 67.4 & 65.3 & 69.5 & 86.2 & 80.7 & 74.3 &          &        \\
\hline
$\lbrace 0,1\rbrace$ vs.\ $\lbrace 2,3\rbrace$ & Both & 81.0 & 79.3 & 82.6 & 82.0 & 80.0 & 80.6 & 40{,}000 & 0\\
\hline
\end{tabular}
\end{table}

The distinction matters because the off-pair predictions are not a random subset of the images. They are concentrated among the harder cases, where the classifier was unsure and assigned an intermediate label rather than committing to one of the two contrasted classes. Conditioning on a within-pair prediction therefore retains
the easier images preferentially, and the conditional accuracy is correspondingly optimistic. For the $0$ versus $3$ contrast the conditional accuracy is 95.9\%, but it is computed only on the 13,968 images the classifier assigned to Normal or Invasive and excludes the 6,032 images it routed to an intermediate class. When those off-pair predictions are counted as errors, the unconditional accuracy for the same contrast falls to 67.0\%.

The unconditional sensitivity and specificity reduce to the per-class recalls of
the four-class classifier. For the $0$ versus $3$ contrast the unconditional
sensitivity for Invasive is 65.3\% and the unconditional specificity for
Normal is 68.6\%, which are exactly the Invasive and Normal recalls in
Table~\ref{tab:svm-class-performance}. The unconditional pairwise summaries
therefore add nothing beyond the four-class diagonal, while the conditional
summaries describe a genuinely different quantity, namely how the classifier
behaves once it has committed to one of the two extremes. We report both and do
not read the conditional accuracy as a measure of how well the method separates
the two classes overall.

Read together, the two treatments give a consistent and clinically sensible
picture. Among the images the classifier assigned to one of the two diagnostic extremes, it almost never reversed them: only 275 true Normal images were called Invasive and only 299 true Invasive images were called Normal, so direct Normal-Invasive confusion is rare. At the same time, the unconditional accuracy shows that a substantial share of Normal and Invasive images are not separated outright, because the classifier resolves its uncertainty by routing them to Benign or InSitu rather than by flipping one extreme to the other. The most frequent errors are assignments to adjacent diagnostic categories, not direct pairwise reversals.

The same pattern holds for the remaining pairwise contrasts, with smaller off-pair counts and correspondingly smaller gaps between the conditional and unconditional accuracies. For Benign versus InSitu the accuracy is 86.4\% conditionally and 70.1\% unconditionally, and for Benign versus Invasive it is 83.3\% conditionally and 67.4\% unconditionally. The grouped normal-or-benign versus carcinoma-type contrast has no off-pair set, and its accuracy is 81.0\%. Because nothing is discarded, this grouped summary is the most directly interpretable binary measure. It indicates that the QNDWT2D descriptors carry a useful malignancy-related signal, while the four-class formulation remains more informative because it shows where the errors occur.

\subsection{Group permutation importance}
\label{subsec:group-importance}
Permutation importance was computed within the held-out folds. For each feature group, the selected columns were permuted as a block using one common random permutation of the test images. This preserves the internal relationships among features in the group while breaking their association with the image labels. The importance score is the resulting decrease in held-out balanced accuracy, averaged over folds, repetitions, and permutation replicates.

\begin{figure}[!htbp]
\centering
\includegraphics[width=0.82\textwidth]{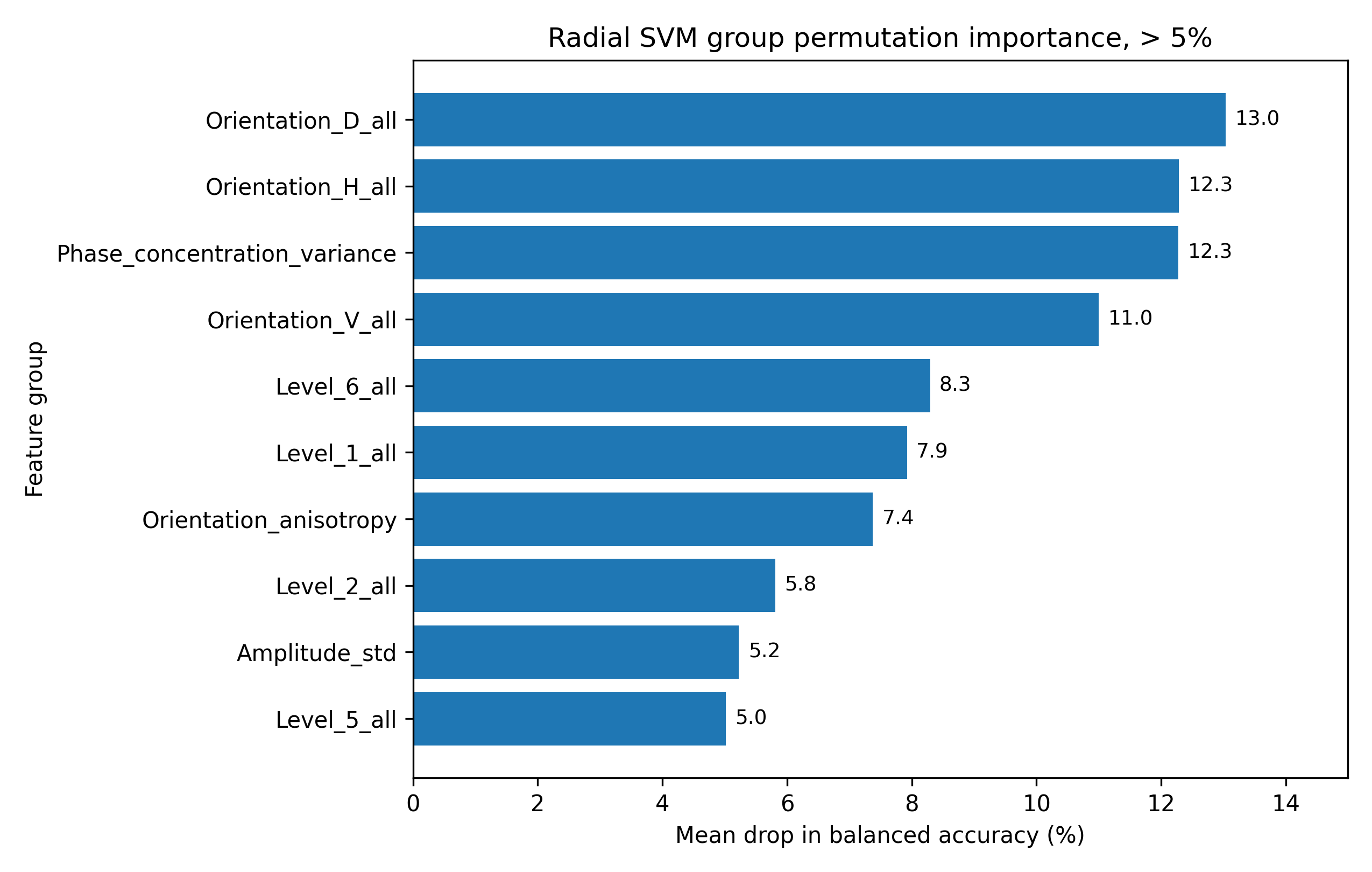}
\caption{\small Group permutation importance for feature classes with mean decrease
in held-out balanced accuracy greater than 5 percentage points. Importance is
computed by block-permuting all features in a group within the held-out folds. The
groups are overlapping rather than disjoint, since a feature can belong to both an
orientation group and a level group, so the bars are marginal drops in accuracy
when each view is destroyed and are not additive contributions.}
\label{fig:group-importance}
\end{figure}

Figure~\ref{fig:group-importance} shows the feature groups whose mean permutation importance exceeds 5 percentage points. The strongest groups are the directional collections: diagonal, horizontal, and vertical features, with mean drops in balanced accuracy of approximately 13.0, 12.3, and 11.0 percentage points, respectively. This is strong evidence that the classifier uses orientation-specific multiscale information rather than relying only on global color summaries or total energy.

The phase concentration and circular variance group is also highly influential, with a mean drop of about 12.3 percentage points. This indicates that the scalar-vector geometry of the quaternion coefficients contributes materially to classification: the fitted SVM benefits not only from coefficient magnitudes but also from summaries of quaternion phase coherence.

The level-wise groups show that both fine and coarse scales matter. Level 6 features produce a drop of about 8.3 percentage points and Level 1 features about 7.9 percentage points, while Level 2 and Level 5 features also exceed the threshold, at about 5.8 and 5.0 percentage points. This supports the multiscale motivation: fine structures such as nuclei and small boundaries are informative, but coarse architecture and broad tissue organization also carry diagnostic information. Orientation anisotropy (about 7.4 percentage points) and amplitude standard deviation (about 5.2 percentage points) further indicate that both directional organization and amplitude variability contribute to discrimination.

\subsection{Individual predictor importance}
\label{subsec:individual-importance}

Individual permutation importance was computed by permuting one predictor at a time in the held-out fold. These individual effects are smaller than the group effects, as expected. The radial SVM uses all standardized predictors through distances in feature space, and many QNDWT2D descriptors are correlated. Therefore, destroying one predictor usually removes only part of the information available to the classifier, because related predictors remain present.

\begin{figure}[htb]
\centering
\includegraphics[width=0.82\textwidth]{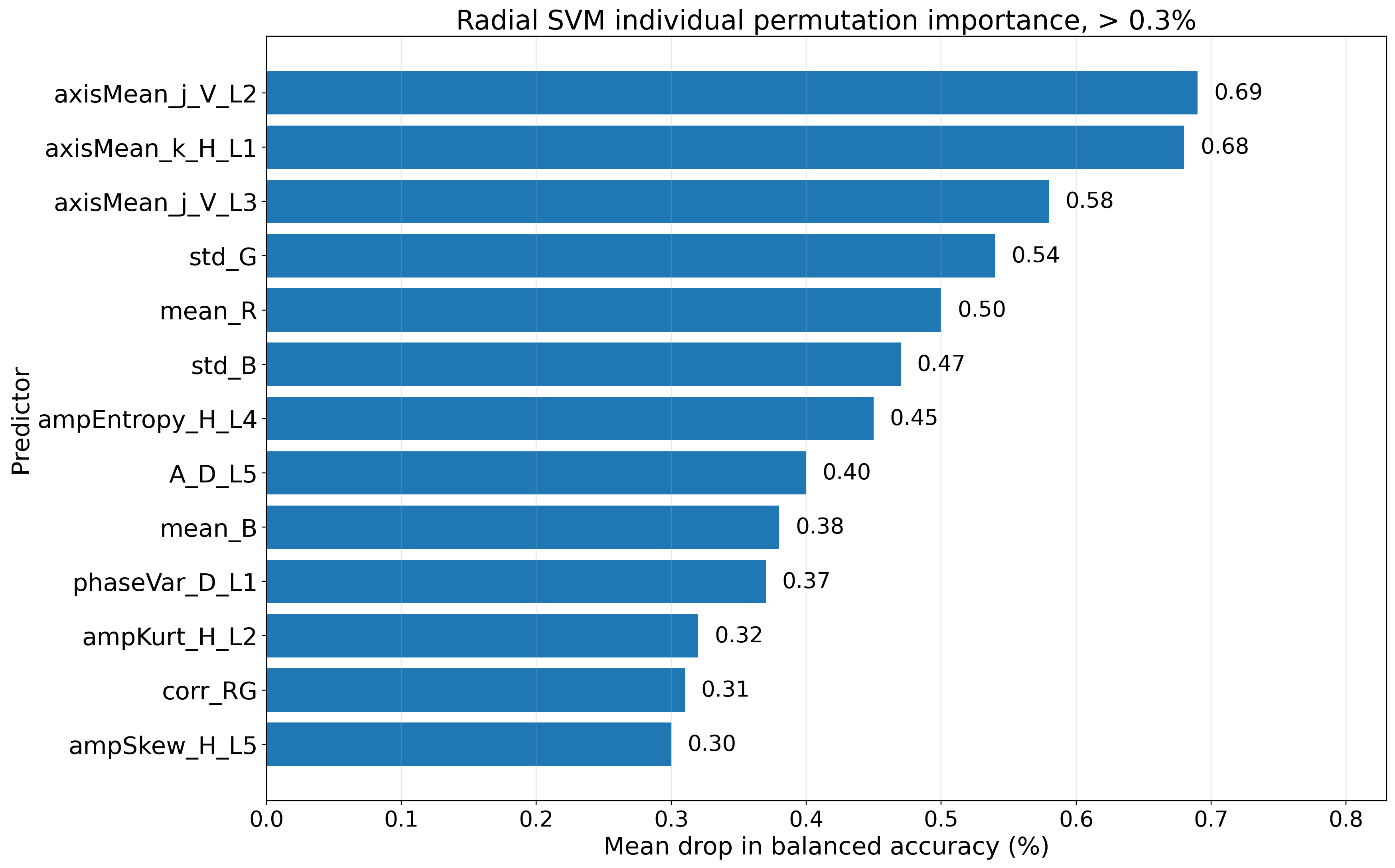}
\caption{\small Individual permutation importance for nonredundant predictors with mean decrease in held-out balanced accuracy greater than 0.3 percentage points. Predictors that were perfectly correlated with stronger predictors were removed from the display. Individual effects are smaller than group effects because the radial SVM uses many correlated predictors jointly.}
\label{fig:individual-importance}
\end{figure}

Figure~\ref{fig:individual-importance} shows the nonredundant individual predictors whose mean drop in held-out balanced accuracy exceeds 0.3 percentage points. Among the 15 predictors that initially exceeded this threshold, two were removed from the final list because they were perfectly correlated with stronger predictors by mathematical construction, so the summary was redundant by design: \feat{p\_D\_L5} was eliminated in favor of \feat{A\_D\_L5}, and \feat{phaseConc\_D\_L1} was eliminated in favor of \feat{phaseVar\_D\_L1}.

The largest remaining individual predictors are quaternion-axis summaries: \feat{axisMean\_j\_V\_L2}, \feat{axisMean\_k\_H\_L1}, and \feat{axisMean\_j\_V\_L3}, with drops of about 0.69, 0.68, and 0.58 percentage points. These variables indicate that directional quaternion-axis behavior in the horizontal and vertical detail bands contributes directly to the radial SVM decision boundary.

Several global color summaries also exceed the 0.3 percentage point threshold. These include the standard deviations of the green and blue channels, the mean red and blue channel intensities, and the red-green channel correlation. This confirms that global stain behavior remains useful even after the wavelet-domain descriptors are included.

\begin{figure}[htb]
\centering
\includegraphics[width=0.90\textwidth]{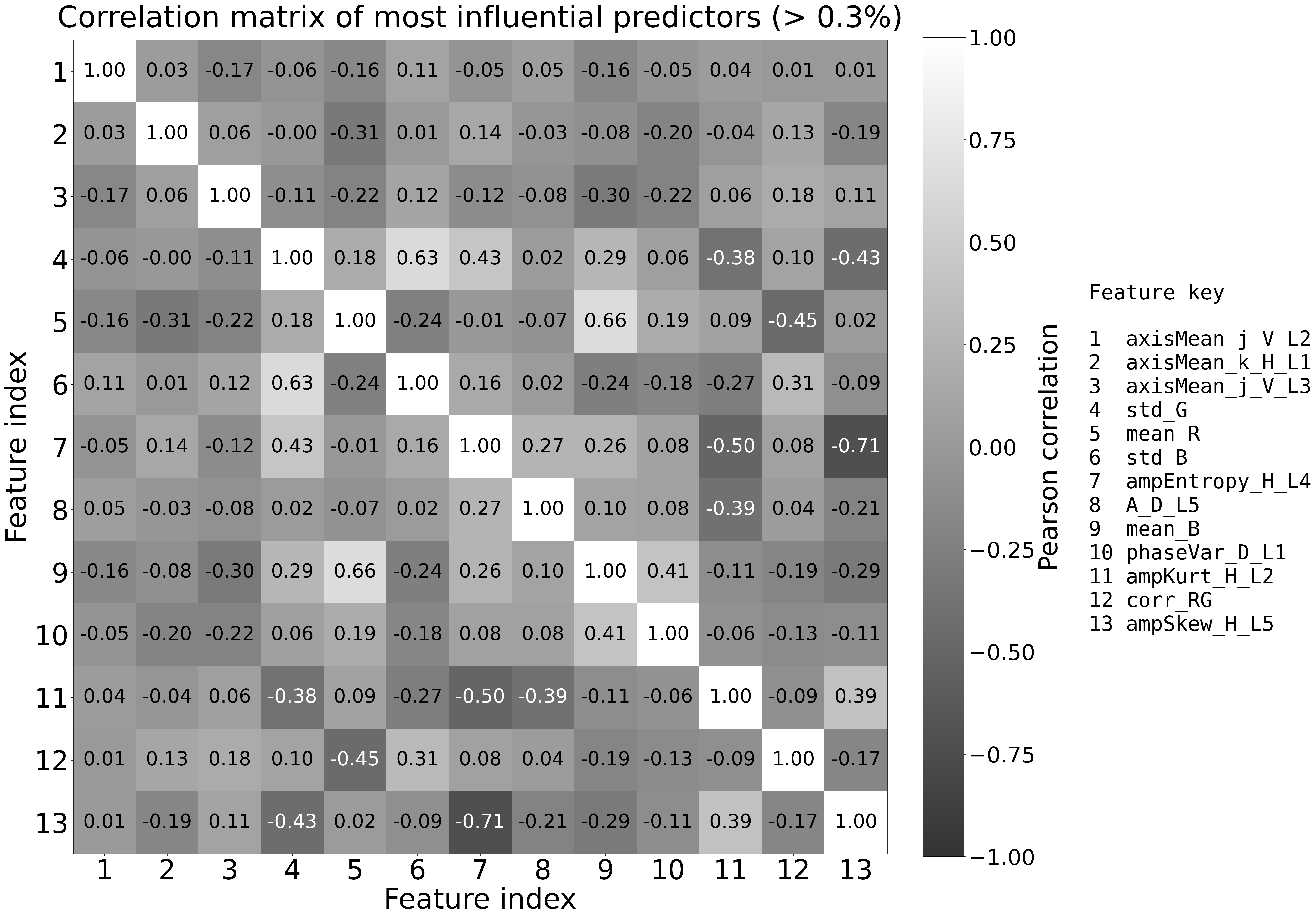}
\caption{Correlation matrix for the nonredundant individual predictors whose mean permutation importance exceeds 0.3 percentage points. The sidebar identifies the predictor names corresponding to the matrix indices.}
\label{fig:individual-importance-correlation}
\end{figure}

The remaining leading QNDWT2D variables include horizontal amplitude entropy at Level 4, diagonal energy share at Level 5, phase variance in the diagonal band at Level 1, horizontal amplitude kurtosis at Level 2, and horizontal amplitude skewness at Level 5. This is a meaningful combination. The fine-scale diagonal phase feature suggests that local oblique or corner-like color-texture responses are useful for discrimination, while the Level 5 diagonal energy-share and horizontal amplitude features indicate that broader architectural content also contributes. The presence of both color marginal features and wavelet-domain phase-orientation features supports the view that the classifier uses both stain information and multiscale tissue geometry. Figure~\ref{fig:individual-importance-correlation} shows the pairwise correlations among these nonredundant predictors.

\section{Conclusions}
\label{sec:conclusions}

This paper proposed a quaternion nondecimated wavelet descriptor for four-class breast histology image classification in the BACH dataset. The central idea was to preserve the original RGB image as a color-vector field, represent each pixel by a pure quaternion, and then extract multiscale, directional, and color-geometric summaries from the resulting quaternion wavelet coefficients. The descriptor was designed before classification, rather than learned from images by pretraining or by large-scale neural optimization. Its purpose is to extract diagnostically interpretable information from the images: summaries of color, texture, scale, orientation, and quaternion coefficient geometry. The aim throughout was a representation whose discriminative power can be read back in histopathological terms, rather than the highest attainable accuracy.

The main classification results show that the QNDWT2D descriptors contain a meaningful four-class signal for the BACH histology problem. The radial SVM achieved an overall accuracy and balanced accuracy of about 68.6\% under repeated cross-validation, with reasonably comparable recognition rates across Normal, Benign, InSitu, and Invasive images. The confusion structure was also clinically plausible. The most difficult errors occurred among adjacent or partially overlapping histopathological categories, especially those involving Benign and InSitu cases, whereas direct confusion between Normal and Invasive images was uncommon. This pattern is important because it suggests that the classifier is not making arbitrary four-class mistakes. Rather, most of its uncertainty follows the expected biological and morphological proximity of the diagnostic classes.

The marginal binary summaries reinforce this reading. When the four-class predictions were restricted or collapsed to clinically meaningful contrasts, the clearest separation was between the diagnostic extremes, and the normal-or-benign versus carcinoma-type contrast showed useful discriminatory ability. These results indicate that the QNDWT2D representation captures a malignancy-related signal, although the full four-class problem remains harder because it requires finer separation among intermediate histopathological states.

Permutation importance was used to interpret the fitted classifier and to identify which extracted features carry diagnostic information. At the group level, the most influential feature classes were not arbitrary. Directional collections, phase concentration and circular variance, level-wise multiscale groups, orientation anisotropy, amplitude variability, and energy and log-energy summaries all contributed substantially to held-out performance. These groups correspond to recognizable histopathological mechanisms: tissue orientation, glandular and stromal organization, nuclear and boundary texture, stain variation, coarse architecture, and multiscale roughness. At the individual predictor level, the most influential nonredundant variables combined quaternion-axis summaries, RGB stain summaries, ...diagonal energy share, diagonal phase variability, and amplitude skewness and kurtosis. Thus the important predictors have direct interpretations in terms of color-vector geometry and multiscale tissue structure.

This interpretability is the central contribution of the approach. Published BACH results based on convolutional neural networks and related deep-learning systems have achieved higher classification accuracy, especially when pretrained architectures, patch-level learning, model ensembles, or extensive tuning were used. The present approach is not intended to compete with those methods on accuracy. Its aim is to extract diagnostically meaningful information from the images through a predetermined handcrafted representation, without pretraining, without learned convolutional filters, and without a deep feature extractor. The resulting features are not latent activations of a neural network. They are mathematically defined summaries of color, amplitude, phase, direction, and scale, and therefore can be discussed in histopathological language.

There are several natural directions for future work. First, the present analysis should be repeated on larger and more heterogeneous histology collections, including images acquired under different scanners, laboratories, staining protocols, and magnifications. This would test whether the QNDWT2D feature families are robust to domain shift. Second, the descriptor could be combined with tissue masks, nuclear segmentation, or glandular region annotations, so that features are computed separately on biologically meaningful compartments rather than over the whole image field. Third, the current image-level summaries could be extended to patch-level and whole-slide settings, where spatial heterogeneity across a slide may itself be diagnostically informative. Fourth, the QNDWT2D features could be used in hybrid models, either as interpretable covariates alongside deep embeddings or as constraints and diagnostic summaries for neural methods. Finally, further work is needed on redundancy reduction and feature stability, since some descriptors are mathematically related by design and others are empirically correlated because they summarize the same underlying coefficient fields. In this sense, the contribution of the paper is not only a classification result, but a reproducible and interpretable quaternion wavelet framework for color histology image analysis.

\subsection*{Data and code}
All programs used to generate the QNDWT2D features, run the radial SVM classification, compute permutation importance, and reproduce the figures have been posted in a public GitHub repository (\url{https://github.com/saraantonijevic/QNDWT2D-histology}) in both MATLAB and Python versions. The repository also includes a guide for reproducing the numerical results from the original image folders through the final tables and figures. This is important for the present study because the feature definitions are explicit and the analysis is intended to be reproducible. A reader can inspect the transform, regenerate the feature table, rerun the classifier, and verify the reported importance rankings.

\bibliographystyle{apalike}
\bibliography{references}

\end{document}